\documentclass[twocolumn,showpacs,superscriptaddress,10pt]{revtex4-1} 
\usepackage{amsmath,amssymb,graphicx,bm,color}
\newcommand\scalefactor{0.8}
\newcommand\scalefactortwo{0.65}
\RequirePackage[normalem]{ulem} 
\RequirePackage{color}\definecolor{RED}{rgb}{1,0,0}\definecolor{BLUE}{rgb}{0,0,1} 

\begin{document}
\title{Composite multi-vortex diffraction-free beams and van Hove singularities in honeycomb lattices}


\author{Vassilis Paltoglou}
\affiliation{Department of Applied Mathematics, University of Crete, 71409 Heraklion, Crete, Greece}
\author{Zhigang Chen}
\affiliation{Department of Physics and Astronomy, San Francisco State University, San Francisco, CA 94132}
\author{Nikolaos K. Efremidis}
\email[Corresponding author: ]{nefrem@uoc.gr}
\affiliation{Department of Applied Mathematics, University of Crete, 71409 Heraklion, Crete, Greece}
\date{}

\begin{abstract}
We find diffraction-free beams for graphene and MoS$_2$-type honeycomb optical lattices. The resulting composite solutions have the form of multi-vortices, with spinor topological charges ($n$, $n\pm1$). Exact solutions for the spinor components are obtained in the Dirac limit. The effects of the valley degree of freedom and the mass are analyzed. 
Passing through the van-Hove singularity the topological structure of the solutions is modified. Exactly at the singularity the diffraction-free beams take the form of strongly localized one-dimensional stripes. 
\end{abstract}


\maketitle



The dynamics of an optical wave can be engineered by either modulating the shape of the wave itself or by changing the properties of the medium. Over the past years a large amount of effort has been devoted in both of these directions. Of central importance is the ability to control the inherent diffraction of propagating waves. One limiting case is to completely suppress diffraction leading to the generation of the so-called diffraction-free beams (DFBs). In homogeneous media there are two main classes of DFBs: The first one is the Bessel beam~\cite{durni-josaa1987,durni-prl1987} which, in terms of ray optics, can be described as an extended and elongated focus. The second class is the Airy beam~\cite{sivil-ol2007,sivil-prl2007} which, 
in terms of ray optics, can be described as a caustic. Along these two directions a variety of different classes of beams have been proposed and observed~\cite{vega-ol2000,bandr-ol2004,mcglo-cp2005,hu-springer2012}. A diffraction-resisting hybrid having the form of the Bessel beam but being able to transversely accelerate along pre-defined trajectories has also been realized~\cite{chrem-ol2012-bessel,zhao-ol2013}. 

The other ingredient which can be appropriately engineered is the medium. Periodic variations of the refractive index can be easily created, using for example optical induction~\cite{efrem-pre2002pr}, resulting to a host of discrete wave phenomena~\cite{leder-pr2008}. 
DFBs can be constructed in square lattices by utilizing the Floquet-Bloch structure of the system~\cite{manel-ol2005}. 
The equivalence between the equations governing the light dynamics in these arrays and the ones in solid state physics or in quantum theories, offers the opportunity to explore in optical settings phenomena predicted in branches of science beyond optics. Perhaps, the most eminent representative of such a structure is graphene, which was recently made available~\cite{novos-science2004}. 
Effort has been made to design graphene-like materials with modified properties.
In MoS$_2$~\cite{radis-nn2011} a minigap opens at the Dirac points, a property highly desirable for potential electronic applications. A bandgap can also open up by deforming a honeycomb lattice~\cite{treid-prl2010,ablow-pra2010,recht-nature2013}.  Relevant phenomena have been studied in honeycomb waveguide arrays including conical diffraction~\cite{peleg-prl2007}, the trembling motion (Zitterbewegung) of electrons~\cite{zhang-prl2008}, the Klein tunneling~\cite{treid-prl2010}, topological insulators~\cite{recht-nature2013} and the relation between pseudospin and orbital angular momentum~\cite{song-nc2015}.

In this work, we find DFBs for graphene and MoS$_2$-type honeycomb optical lattices. We introduce a physically relevant spinor field decomposition method to analyze the underlying structure of these waves. The resulting composite solutions have the form of multi-vortices or semi-vortices, in the sense that the two spinor components are associated with different topological charges ($n$, $n\pm1$). Asymptotically, close to the Dirac point, the system is described by the Dirac-Weyl equation (massless or massive). In this limit, we analytically obtain diffraction-free composite multi-vortex solutions which are in agreement with our numerical results. Furthermore, the effect of the valley inequivalence of the $K$ and $K'$ points leads to vorticies with different topological charges. As the propagation constant traverses the band structure (BS), the diffraction-free solutions undergo a transition exactly at the van-Hove singularity. As a result, after this transition, the spinor components of the DFBs have the same vorticity. Exactly at the van-Hove singularity three different solutions are obtained which have the form of an array of infinite extent in one direction but strongly localized in the orthogonal direction. We also discuss extensions of this work to generic lattices with more than one ``atoms'' per unit cell.

\begin{figure}[t]
\centering
\includegraphics[clip=true,width=\scalefactortwo\columnwidth]{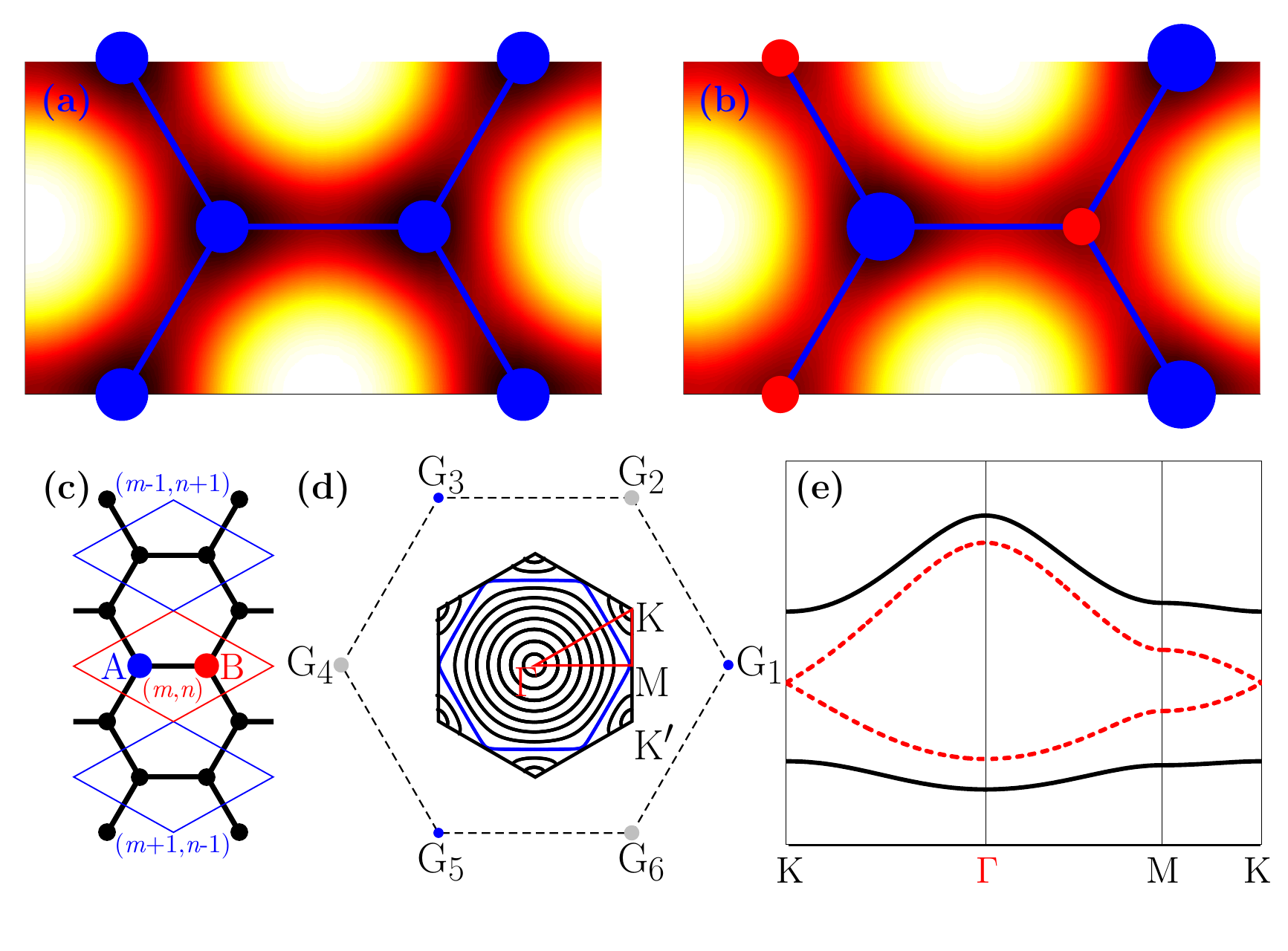}
\caption{Typical index profiles of a graphene (a) and a MoS$_2$ (b) lattice (darker areas represent higher index); (c) lattice structure; (d) contour plot of the graphene BS, (blue curve is the van-Hove singularity);  (e) BS for graphene (dashed red) and MoS$_2$ (solid black) lattices.}
\label{fig:0}
\end{figure}

We consider the paraxial dynamics of light waves in normalized coordinates
$i \psi_z +(1/2)\nabla_\perp^2\psi + V(\bm r) \psi = 0,$
where $\psi$ is the field amplitude, $\bm r=(x,y)$ are the transverse coordinates and $z$ is the propagation coordinate, $\nabla_\perp^2=\partial_x^2+\partial_y^2$ is the diffraction operator and $V(\bm r)$ is the potential that is proportional to the refractive index contrast. In particular, we select a honeycomb periodic potential of the form
$
V(\bm r)=
(V_0/c_0^2) |
\sum\nolimits_{j=1}^6 e^{i (\bm G_j\cdot \bm r +\phi_j)}
-c_0 |^2,
$
where $\bm G_1 = G \hat{\bm x}$, $G=2\pi/L$, $\bm G_j=R^{j-1}(\pi /3)\bm G_1$, $L$ is the lattice period, and $R$ the rotation matrix and $\phi_{2j+1}=\chi/2$, $\phi_{2j}=-\chi/2$. In the rest of the paper we choose $c_0=6$ and $\chi = 0$ for  graphene lattices and $\chi=\pi/20$ for optical MoS$_2$ lattices (honeycomb with sublattice symmetry breaking) [see Fig.~\ref{fig:0}]. Note that similar lattices can be generated via optical induction as shown 
in~\cite{gao-ieeepj2014}. 
Due to symmetry breaking, in MoS$_2$ lattices a bandgap opens up at the $K$ and $K'$ points of the Brillouin zone [Fig.~\ref{fig:0}(e)].


For relatively high index constrasts, we can apply coupled mode theory (CMT).
We denote the lowest eigenmode of an isolated waveguide of the potential by $U^j_{m,n}(x,y)$, where the subscripts $(m,n)$ label the translations over the unit cell [Fig.~\ref{fig:0}(c)], while the superscript $j=\{A,B\}=\{1,2\}$ denotes the two sites in each unit cell. We decompose the optical field as $\psi=\sum_{m,n,j}c_{m,n}^j(z)U^j_{m,n}(\bm r)$, where $c^j_{m,n}$ are the $z$-varying amplitudes which can be written in spinor form as $c_{m,n}=(c^A_{m,n},c^B_{m,n})$. 
Following the relevant calculations we obtain 
$
i\frac d{dz}c_{m,n}^j+t\sum_{\langle m,n\rangle}c_{m,n}^{3-j}+(-1)^{j+1}\beta_0c_{m,n}^j=0
$
where $t$ is the coupling strength between adjacent waveguides, $\langle m,n\rangle$ denotes coupling over first neighbors, and $\pm\beta_0$ are the relative shifts of the propagation constants due to the sublattice asymmetry resulting to a bandgap width $2\beta_0$. 

\begin{figure}[t!]
\centering
\includegraphics[clip=true,width=\scalefactor\columnwidth]{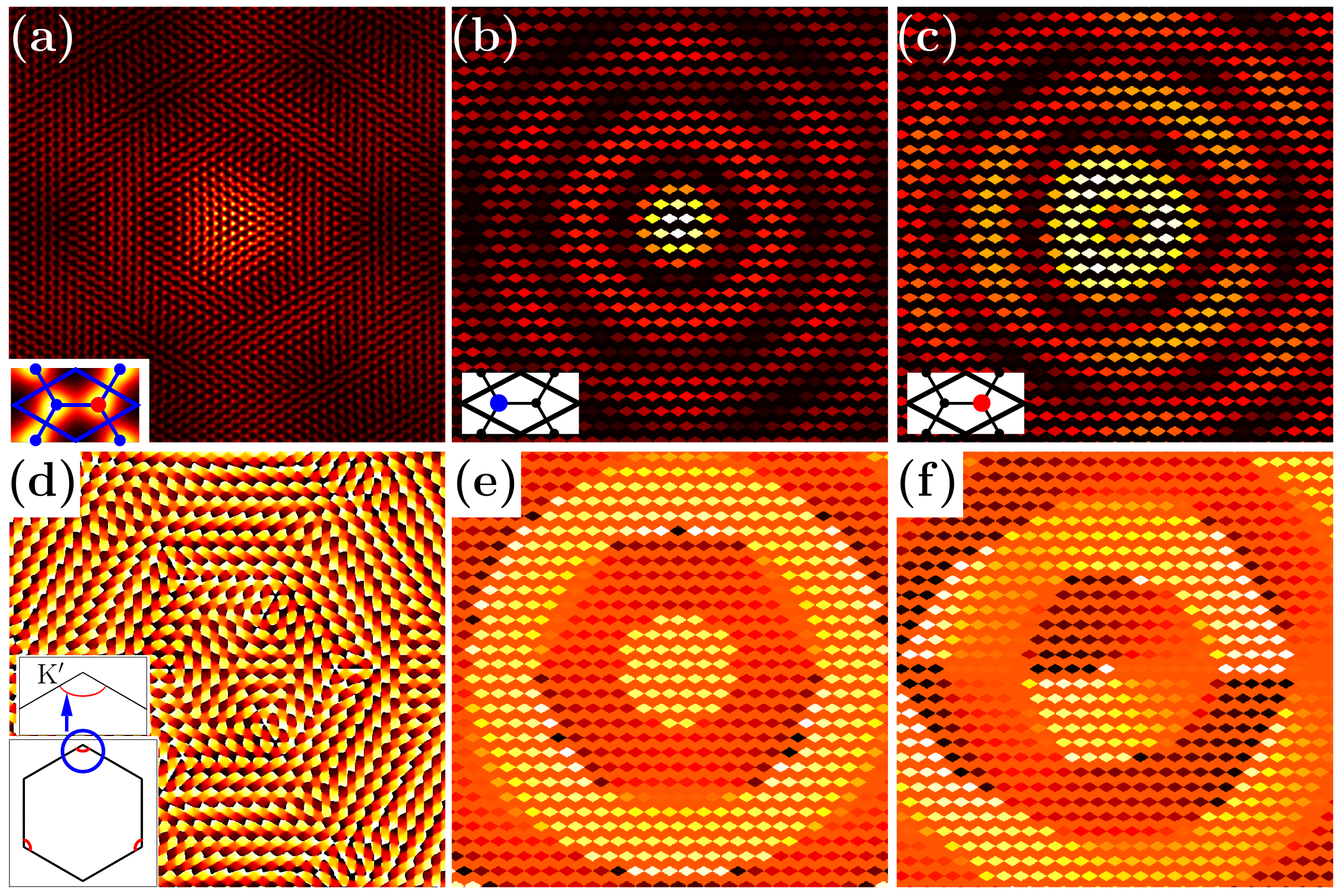}
\caption{A diffraction-free composite semi-vortex with topological charges $Q=(0,-1)$ supported by a graphene type lattice for $V_0=15$ and $L=1.81$. The propagation constant is $\beta = -19.099$ and the Dirac point is $\beta_D=-18.905$. In the first and second rows the intensity and phase patterns are shown. In column (a) the total field $\psi$ is shown, while in columns (b) and (c) the isolated spinor components $\psi_A$ and $\psi_B$ are presented (the insets show the corresponding sublattices).  The inset of  (a) is a typical Bloch mode amplitude along the trajectory  shown in inset (d).}
	\label{fig:1}
\end{figure}

The continuous limit of the CMT equations can be derived [$c_{m,n}(z)\rightarrow u(x,y,z)=(a,b)$] by selectively exciting the lattice with a broad beam in the vicinity of particular $\bm k$-vectors of the BS~\cite{ablow-pra2009}. Specifically, near the vertices ($K$ and $K'$) of the 1st Brillouin zone of Fig.~\ref{fig:0}(e) the following Dirac equation is derived
$i \partial_z a - D( \partial_\mp b) + \beta_0 a  =  0 $
$i \partial_z b + D( \partial_\pm a) - \beta_0 b  =  0,$
where $D=\sqrt{3} tL /2$ and $\beta_0$ describes the effective mass in the context of MoS$_2$ and is zero for graphene. In addition  
$\partial_\pm = \partial_x \pm i \partial y = e^{\pm i \phi}\left( \partial_r \pm i \frac{1}{r} \partial_\phi
\right)$ and the signs refer to the two inequivalent Dirac points (upper sign is $K$ and lower sign is $K'$). Without loss of generality, we set $D=1$ for the rest of the paper. 

We have found DFB solutions of the Dirac system which are expressed in terms of Bessel functions as
\begin{align}
a(r, \phi, z) & =  J_n(\lambda r) e^{i n \phi} e^{-i \beta z},
\label{eq:dirac1}
 \\
b(r, \phi, z) & =  g J_{n\pm 1}(\lambda r)e^{i(n\pm 1)\phi}e^{-i \beta z} ,
\label{eq:dirac2}
\end{align}
where 
$g(\beta)=\mp\sqrt{(\beta+\beta_0)/(\beta-\beta_0)}$, and $\lambda = \sqrt{\beta^2-\beta_0^2}$. Thus the spinor $u=(a,b)$ consists of two vortices with unequal topological charges $Q=(n,n\pm1)$. We call this DFB a \textit{composite multi-vortex} DFB. We note that for graphene lattices $\beta_0=0$ and thus the two spinor components are equally ``inhabited'' ($|g|=1$). On the other hand, the presence of an effective mass $\beta_0$ gives rise to a completely different behavior. As we approach the band edges $\beta\rightarrow\beta_0$ ($\beta\rightarrow-\beta_0$) the relative amplitude of $b$ ($a$) goes to zero and we end up with a single vortex with vorticity $n$ ($n\pm1$), respectively. As we move away from the gap edges, the relative amplitude of the two components become equal. 

\begin{figure}[t!]
  \centering
  \includegraphics[clip=true,width=\scalefactor\columnwidth]{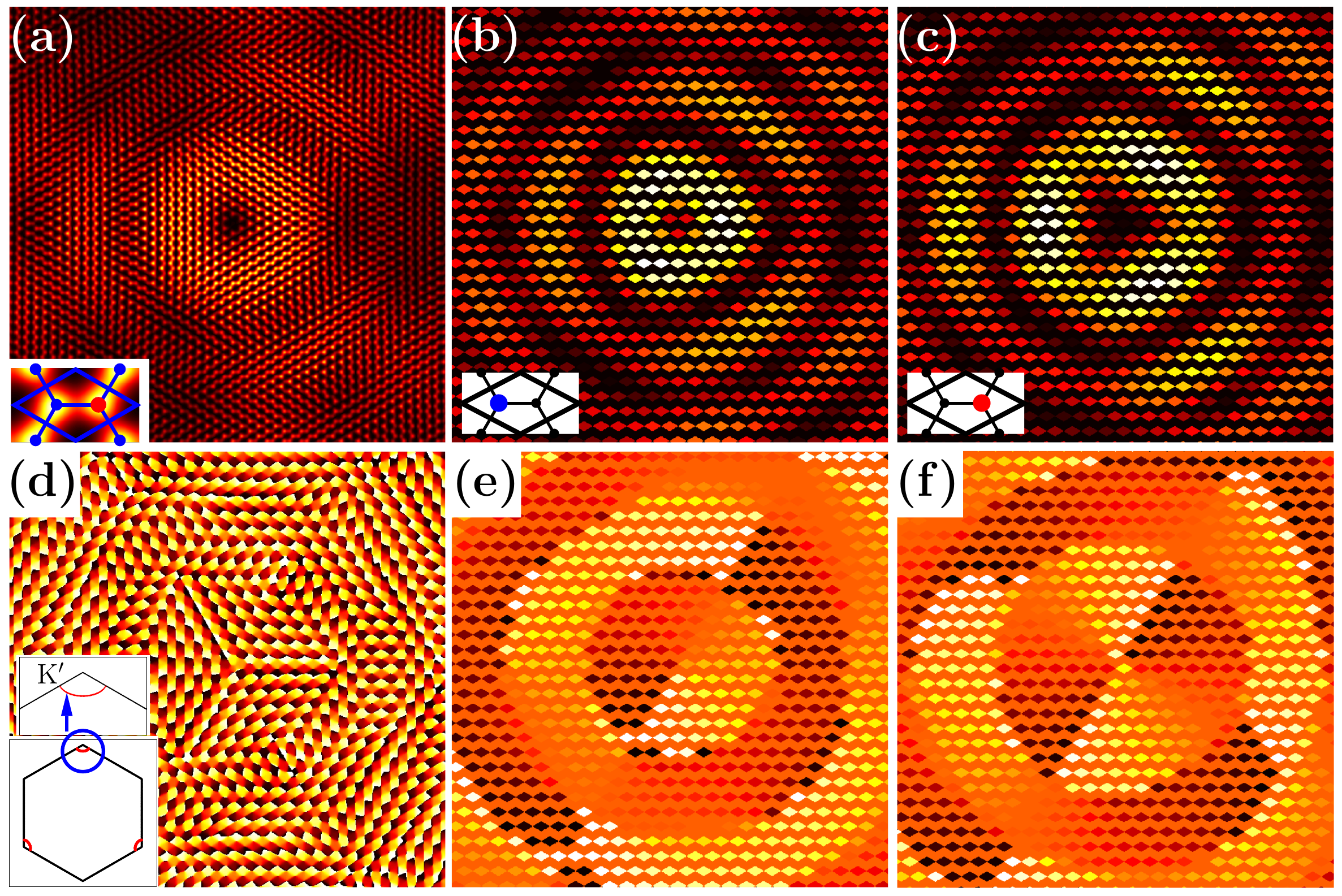}
  \caption{Same as in Fig.~1 for a diffraction-free composite multi-vortex with topological charges $Q=(-1,-2)$.}
  \label{fig:2}
\end{figure}
It is interesting to point out that, in contrast to the Dirac equations, within the paraxial framework both spinor components are described by a single wavefunction~\cite{song-nc2015}. Thus, it is physically relevant to follow a similar decomposition, for the paraxial model. In particular we separate the wavefunction of the paraxial equation  discretely according to its spatial location as $\psi_A(\bm R_{m,n}^A,z)$ and $\psi_B(\bm R_{m,n}^B,z)$. In the above equations $\bm R_{m,n}^j$  are the position vectors of the lattice elements $A$ and $B$  [see Fig.~\ref{fig:0}(c)] located in the same Wigner-Seitz cell $(m,n)$.  Consequently, there is absolutely no meaning in any form of interference between these two components of the Dirac equation (at least when the CMT is applicable).

\begin{figure}[t!]
\centering
\includegraphics[clip=true,width=\scalefactor\columnwidth]{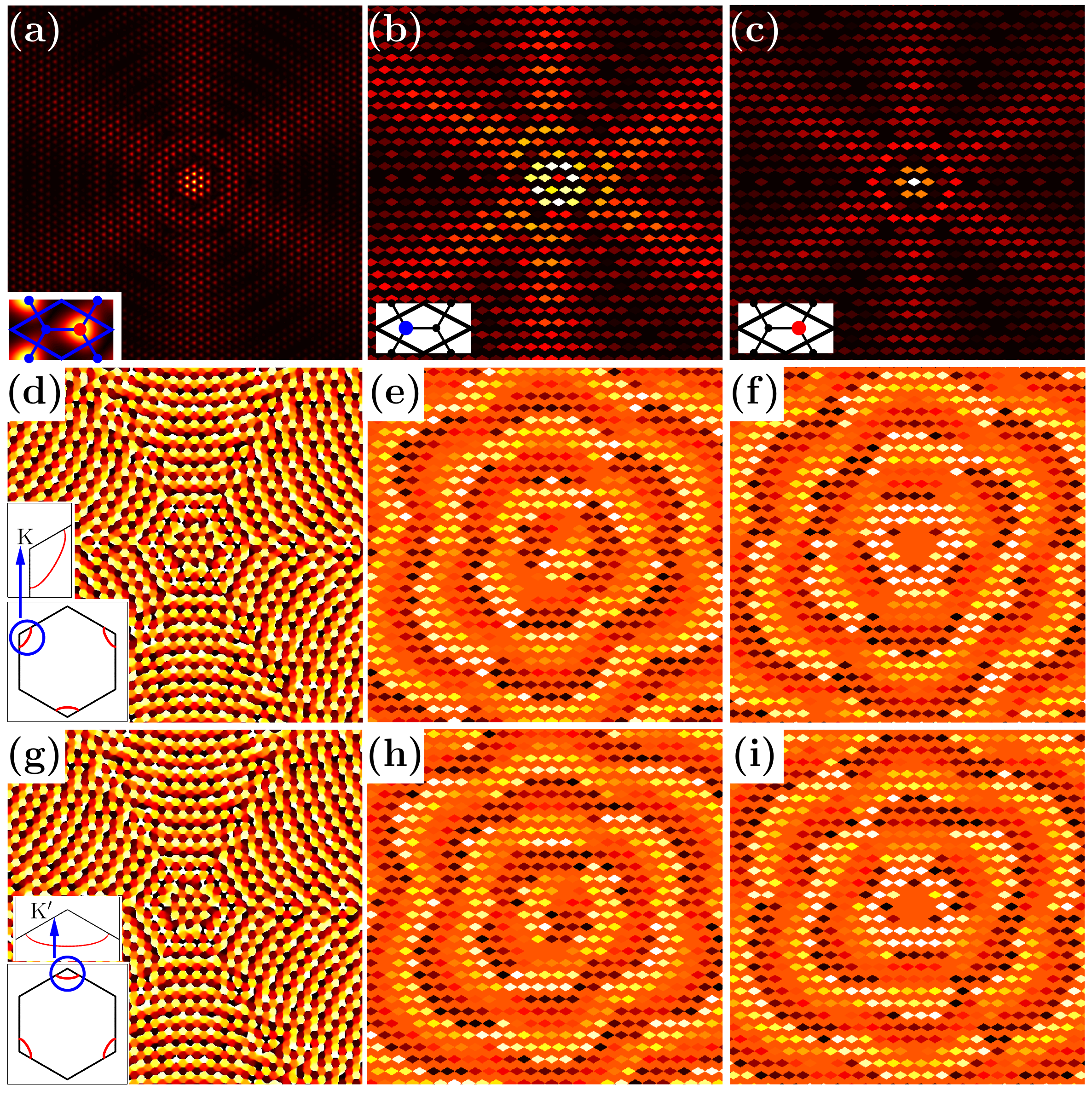}
\caption{Valley and mass effects of MoS$_2$ composite semi-vorticies for $ \beta = -20.424$, $V_0=15$, and $L=1.81$ encircling the $K$ ($K'$) point in second (third) row leading to $Q=(-1,0)$ [$Q=(1,0)$], respectively. The intensity pattern shown in the first row is identical in both cases. }
\label{fig:34}
\end{figure}
The paraxial equation supports Floquet-Bloch modes having the form
$u_{\bm k}(\bm r) = v_{\bm k}(\bm r)e^{i\bm k\cdot\bm r-i\beta(\bm k) z}$
where $v_{\bm k}(\bm r)$ is a function with the same periodicity as the lattice. 
However, if $u_{\bm k}$ is a Floquet mode the same holds for $u_{\bm k}e^{i\xi}$ for an arbitrary phase $\xi$. In the case of a single Floquet mode, this might seem a trivial generalization, however, when superimposing Floquet modes their relevant phase becomes important. We make two different choices for this phase factor and denote the corresponding Floquet modes as $u_{\bm k}^j$, $j={\{A,B\}}$. In particular, we choose all the modes to have a positive real amplitude on the sublattice with $(m,n)=(0,0)$ ($u_{\bm k}^{A}(\bm R^{A}_{0,0})>0$ and $u_{\bm k}^{B}(\bm R^{B}_{0,0})>0$). 

The function $\beta(\bm k)$ describes the BS surface. As we can see in Fig.~\ref{fig:0}(d) the isocontours ($\beta=\mathrm{constant}$) surround the $K$ and $K'$ or the $\Gamma$ points of the BS and, as a result, can be described in terms of a polar coordinate $\varphi$. The separating curve corresponds to the van-Hove singularity. We can invert the relation $\beta(\bm k)=\beta_0$ and solve in terms of the Bloch momentum as $\bm k(\varphi,\beta_0)$ for each curve. We conclude that the general form of a DFB is 
$
\psi = \int_0^{2\pi}A(\varphi)e^{i\Theta(\varphi)}
u_{\bm k(\varphi)}^j e^{i\bm k(\varphi)\cdot\bm r-i\beta(\bm k) z}
d\varphi
$
for arbitrary values of the amplitude $A(\varphi)$ and the phase $\Theta(\varphi)$. Note, that, as we described, for the same value of $\beta$ surrounding the Dirac cones there are four different families of diffraction free beams depending whether we encircle $K$ or $K'$ and on the Floquet modes $u_{\bm k}^A$ or $u_{\bm k}^B$.

In Fig.~\ref{fig:1} we show a DFB that is generated by equal ($A(\varphi)=1$) in-phase ($\Theta(\varphi)=1$) excitation of the Floquet modes $u_{\bm k}^A$ along the, almost circular, 1st band closed contour that surrounds the $K'$ points. Thus, we expect that the asymptotic solutions [Eqs.~(\ref{eq:dirac1})-(\ref{eq:dirac2})] are in good agreement with the numerical solutions. The underlying structure of the wave is revealed by decomposing the field to its spinor components as we previously described. Numerically, in each unit cell $(m, n)$ we calculate the optical field at the index maxima of the sites $A$ and $B$. We then set the value of the field everywhere inside this cell to be equal to this latter calculated value. Since the Floquet modes around the Dirac cone consist of a periodic array of vortices, it is relevant to keep only one out of three unit cells to avoid the imaging of this local vortex structure. Using this decomposition, we clearly see the formation of a composite semi-vortex with topological charges $Q=(0,-1)$ in agreement with Eqs.~(\ref{eq:dirac1})-(\ref{eq:dirac2}) with $n=0$ for the $K'$. In Fig.~\ref{fig:2} we choose the same parameters as in Fig.~\ref{fig:1} but we impose a vortex phase structure $\Theta(\varphi)=-2\pi l(\varphi)/L$, where $l(\varphi)$ is the arclength. This leads to a composite multi-vortex DFB with $Q=(-1,-2)$ [see Eqs.~(\ref{eq:dirac1})-(\ref{eq:dirac2}) with $n=-1$]. 

\begin{figure}[t!]
\centering
\includegraphics[clip=true,width=\scalefactortwo\columnwidth]{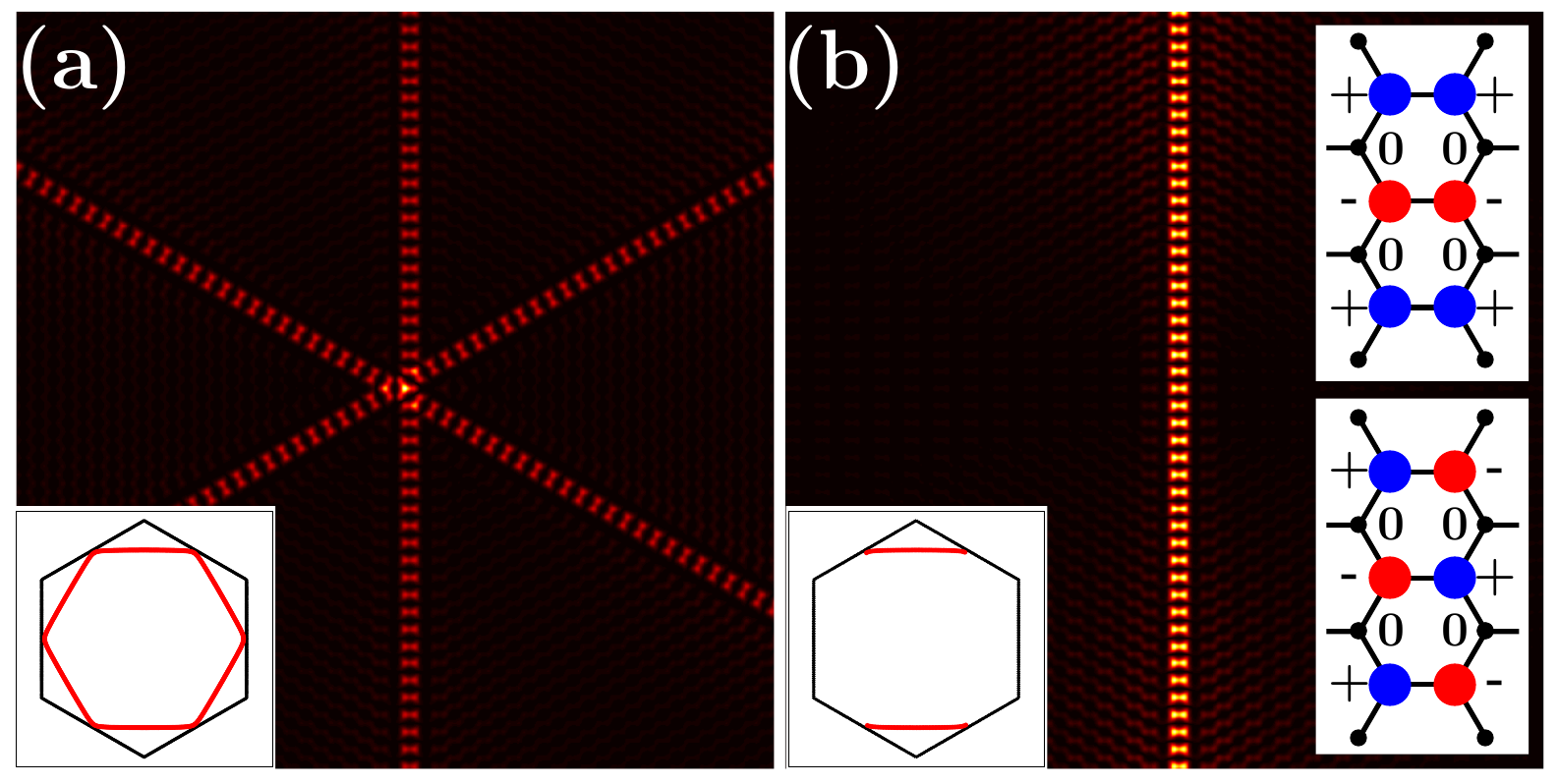}
\caption{Directional 1st band DFB supported by the van-Hove singularity when traversing (a) the closed hexagonal path and (b) two parallel line segments. The structure of the modes is shown in the right insets of (b). A graphene lattice is chosen with $V_0=15$ and $L=1.81$, while $ \beta = -19.442$.}
\label{fig:5}
\end{figure}
In Fig.~\ref{fig:34} we analyze the effects of (i) the valley degree of freedom and (ii) the mass on the formation of DFBs. We select 1st band contours with the same $\beta$ encircling the inequivalent $K$ and $K'$ points. A typical Floquet-mode beam intensity is localized on sublattice $B$ [Fig.~\ref{fig:34}(a) inset] although the index is higher on sublattice $A$ [Fig.~\ref{fig:0}(b)]. This happens due to the local vortex structure that enhances the energy of the respective Hamiltonian. 
Thus, we choose the Floquet modes $u_{\bm k}^B$ which constructively interfere on sublattice $B$ and $A(\varphi)=1$, $\Theta(\varphi)=0$. Due to the valley degree of freedom (inequivalence of $K$ and $K'$) the generated composite semi-vortices, although they have the same intensity pattern, are associated with charges $Q=(-1,0)$ and $Q=(1,0)$, as also predicted from Eqs.~(\ref{eq:dirac1})-(\ref{eq:dirac2}). The effect of the mass leads to another interesting feature. Specifically, the resulting composite DFB has strongly unequal amplitudes in its spinor components as we approach the band edges. In Fig.~\ref{fig:34} the relative maximum amplitude of the spinor components is $\max |\psi^A|/\max |\psi^B|=0.06$.

\begin{figure}[tb]
\centering
\includegraphics[clip=true,width=0.6\columnwidth]{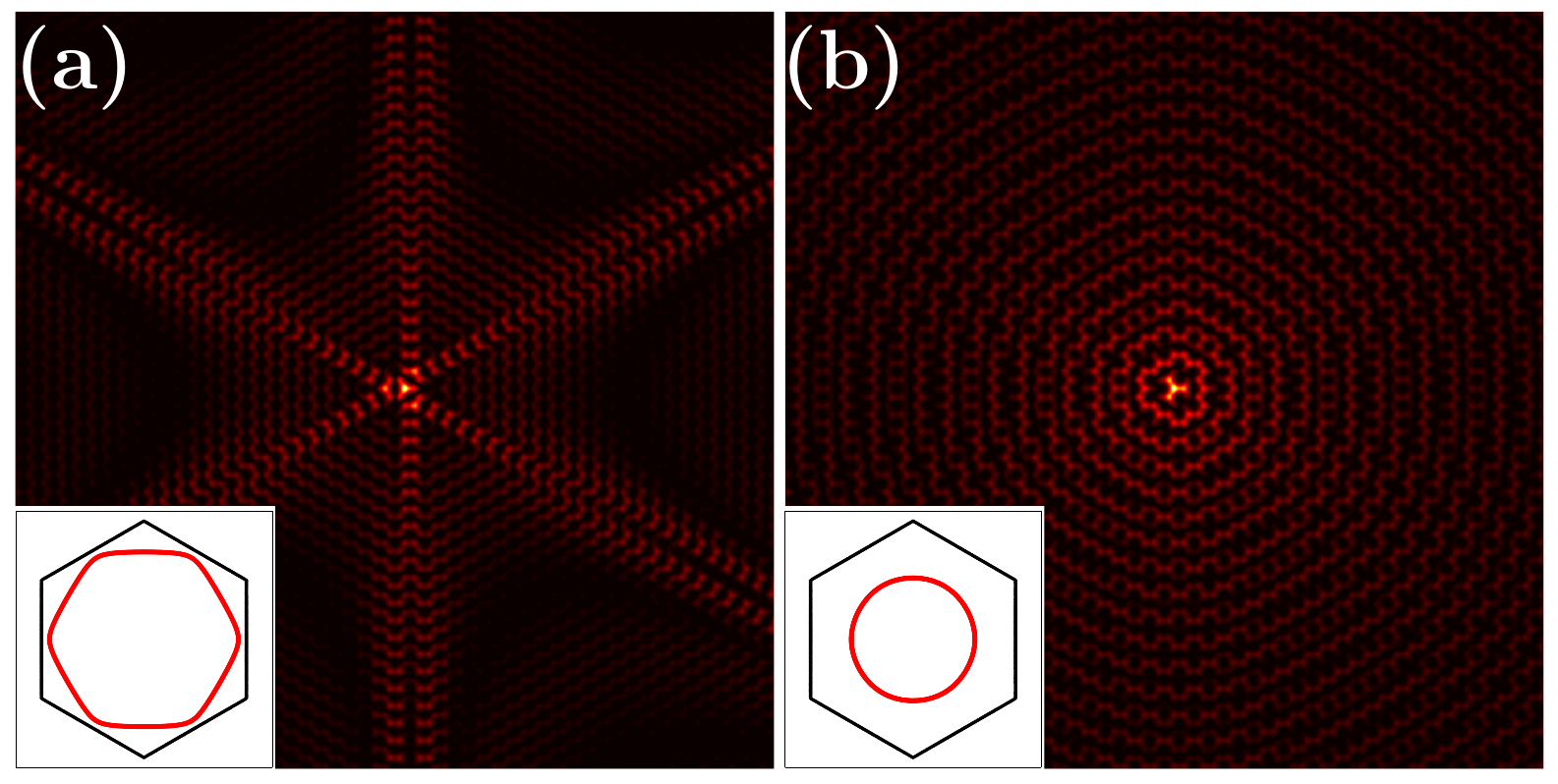}
\caption{DFB encircling the $\Gamma$-point for (a) $\beta = -19.468$ and (b) $\beta = -19.784$ in a graphene lattice with $V_0=15$ and $L=1.81$.}
\label{fig:6}
\end{figure}
As we move further away from the Dirac points the solutions remain topologically equivalent to those presented above, having the same topological charges $Q$. Eventually, we reach the van-Hove singularity which is the contour that passes through the boundary midpoints of the hexagonal 1st Brillouin zone. In Fig.~\ref{fig:5}(a) we equally excite the path that constitutes the van-Hove singularity. As we can see the DFB is associated with three dominant directions. The structure of the DFBs at the singularity is revealed by selectively exciting only two parallel lines segments. In Fig.~\ref{fig:5}(b) we clearly see a DFB that is of infinite extent in the $y$-direction and strongly confined along the $x$-direction. Three such solutions exist with a $2\pi/3$ rotational symmetry. Within the framework of the CMT these solutions are expressed analytically as $c_{l,-l}=(-1)^l\kappa(1,-(\beta_0+\beta_\pm)/t)e^{-i\beta z}$ and $c_{m,n}=0$ otherwise, where $\beta_\pm=\pm\sqrt{\beta_0^2+t^2}$ [the indices are shown in Fig.~\ref{fig:0}(c)]. The $\beta_-$ ($\beta_+$) eigenvalue lies in the first (second) zone of the BS [see top (bottom) right inset of Fig.~\ref{fig:5}(b)]. In the graphene limit ($\beta_0=0$) the nonzero amplitudes of the modes in the two sublattices become equal. In MoS$_2$ lattices when $\beta_0/t\gg1$ the amplitudes in one sublattice become negligible resulting to a single out-of-phase straight array of diffraction-free light beams.  

The van-Hove signularity separates the BS into topologically inequivalent regions. Passing through this singularity we approach the neighborhood of the $\Gamma$ point of the Brillouin zone (which the DFB encircles) and the Floquet modes lose the local vortex structure. Typical DFBs in this case are shown in Fig.~\ref{fig:6}.  Specifically, in Fig.~\ref{fig:6}(a) the path is close to the van-Hove singularity and exhibits a ``starfish'' like structure which is reminiscent of the directional modes shown in Fig.~\ref{fig:5}. Moving towards the $\Gamma$ point the integration path as well as the DFB take a cicular form [Fig.~\ref{fig:6}(b)].

The analysis presented above is not limited to honeycomb lattices but can be applied to any lattice with more than one ``atom'' per unit cell. Our investigations show that the presence of the van-Hove singularity separates regions where the composite DFBs have different topological charges. However, in many lattices the van-Hove singularity exists exactly at the band edge. For example, in the case of a square diatomic lattice (where the singularity exists in the band edge) both spinor components have the same vorticity everywhere inside the band. Also, exactly at the singularity elongated DFBs always exist. 

Finally, we would like to point out that DFBs not only constitute a fundamental class of solutions and the building blocks for other classes of beams but are also important in unveiling the fundamental properties of the system. For example, conical diffraction can be explained by expanding the initial condition to the diffraction-free beams supported by the system~\cite{song-nc2015}.


Supported by the action ``ARISTEIA'' (Grant No. 1261) in the context of the Operational Programme ``Education and Lifelong Learning'' that is cofunded by the European Social Fund and National Resources and by US AFOSR and NSF.


\newcommand{\noopsort[1]}{} \newcommand{\singleletter}[1]{#1}

\cleardoublepage

{\bf References with titles}

\end{document}